\documentclass[preprint,12pt]{elsarticle}

\usepackage[english]{babel}
\usepackage[utf8x]{inputenc}
\usepackage[T1]{fontenc}
\usepackage{amssymb}

\usepackage[a4paper,top=3cm,bottom=2cm,left=3cm,right=3cm,marginparwidth=1.75cm]{geometry}

\usepackage{amsmath}
\usepackage{graphicx}
\usepackage[colorinlistoftodos]{todonotes}
\usepackage[colorlinks=true, allcolors=blue]{hyperref}
\usepackage{lineno}

\usepackage{hyperref}

\begin{document}
\begin{frontmatter}
\title{Development of a technology for the fabrication of Low-Gain Avalanche Diodes at BNL}

\author{Gabriele Giacomini\corref{cor1}}
\ead{giacomini@bnl.gov}
\cortext[cor1]{Corresponding author}
\author{Wei Chen\corref{cor2}}
\author{Francesco Lanni\corref{cor3}}
\author{Alessandro Tricoli\corref{cor4}}
\address{Brookhaven National Laboratory, Upton 11973, NY, U.S.A.}

\begin{abstract}
Low-Gain Avalanche Detectors are gathering interest in the High-Energy Physics community thanks to their fast-timing and radiation-hardness properties. One example of this includes plans to exploit timing detectors for the upgrades of the ATLAS and CMS detectors at the High Luminosity LHC.  
This new technology has also raised interest for its possible application for photon detection in medical physics, imaging and photon science.
The main characteristic of this type of device is a thin and highly-doped layer that provides internal and moderate gain, in the order of 10-20, that enhances the signal amplitude. Furthermore the thin substrate, of only few tens of microns, allows for fast carrier collection. 
This paper offers details on the fabrication technology, specifically developed at Brookhaven National Laboratory for the detection of minimum ionizing particles. The static electrical characterization and the gain measurements on prototypes will be  also reported.  
\end{abstract}


\end{frontmatter}


\section{Introduction}

Studies of fast-timing detectors have attracted widespread interest in the scientific community around the world in relation to the development of the next generation of collider experiments and imaging techniques for a variety of applications ~\cite{Sadrozinski:2013nja}. Silicon-based particle sensors that combine fast-timing and radiation-hardness properties are a rapidly developing research area, thanks to their applications in timing and tracking systems for the upgraded ATLAS~\cite{Collaboration:2623663} and CMS~\cite{Collaboration:2296612} experiments at the CERN Large Hadron Collider (LHC) for the High Luminosity phase (HL-LHC), which is expected to begin in 2026~\cite{CERN-ATS-2012-236,CERN-ACC-2015-0140}.
One of the selected technologies for such detectors, that can deliver the  required timing performance of a few tens of ps, is a silicon pad with internal amplification: the so called Low-Gain Avalanche Diode (LGAD)~\cite{Fernandez-Martinez:2015bda}. 

\begin{figure}
\centering
\includegraphics[width=0.9\textwidth]{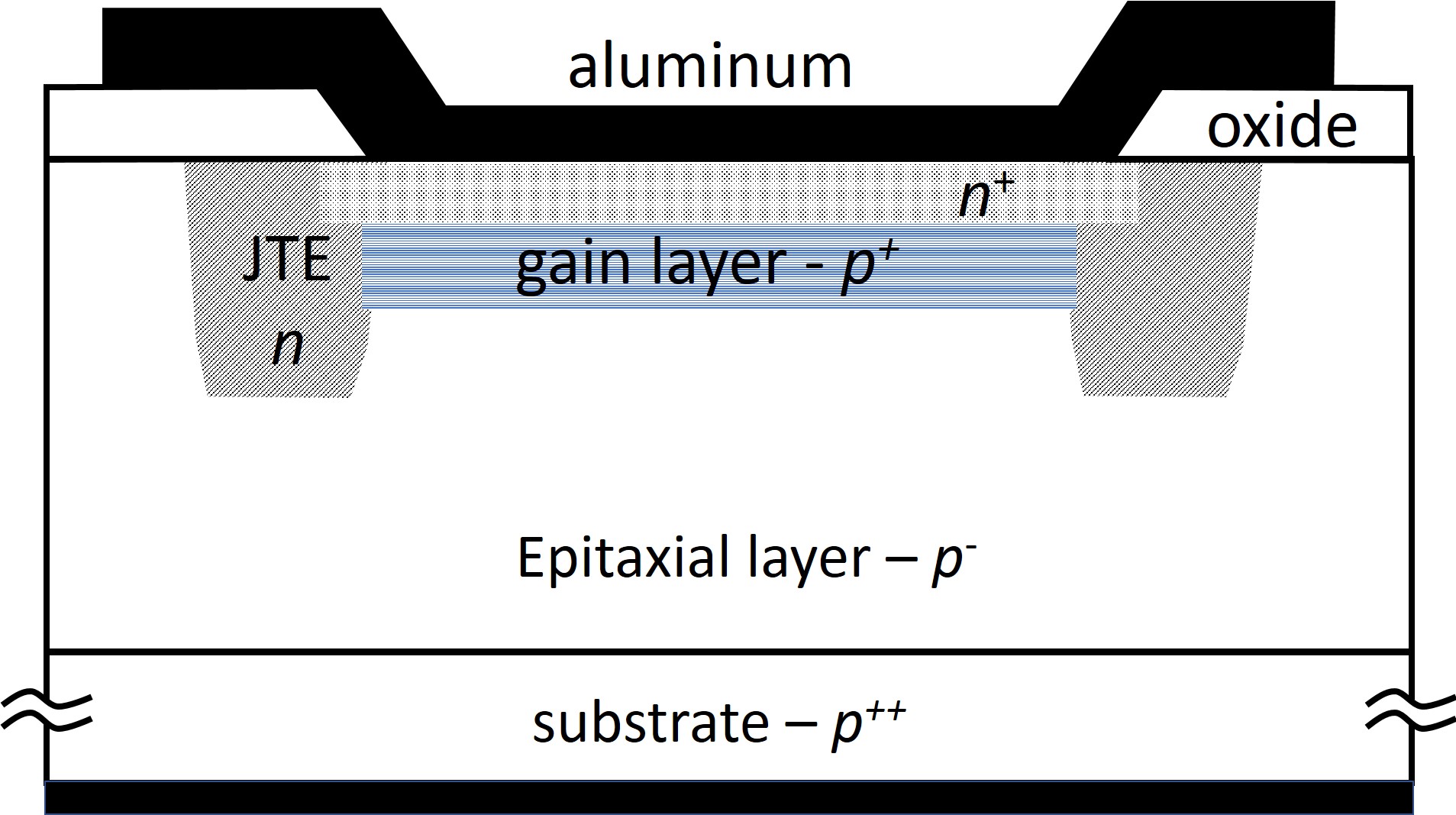}
\caption{\label{fig:sketch} Sketch of the structure of an LGAD (not to scale).  The GR termination is not shown.}
\end{figure}

The structure of a typical LGAD is shown in Figure~\ref{fig:sketch}. The active  volume  of the device is an epitaxial layer or a thin silicon  wafer-bonded to a thick substrate that acts as a mechanical support. The thickness of the epitaxial layer can be a few tens of $\mu$m in order to have a short drift of the carriers. An $n^+ -$layer creates the junction with the $p -$type substrate, as in a regular diode. Additionally, a  $p-$layer is implanted just below the $n^+$ implant ({\it gain layer}). Application of a bias to the junction leads to depletion of this layer which results in a high electric field in a superficial region that extends in depth for about a micron.
 Electron impact ionization is generated by this high field when the drifting electrons enter the gain layer volume,
while  the ionization rate of the holes is at a negligible level (thus excluding  the onset of a breakdown). The result is an amplified current pulse which is  dominated by the motion of the holes through the whole thickness of the substrate. These current pulses, amplified  by a factor typically in the range 10-20, are inherently fast and can offset the limited amount of charge released by a minimum ionizing particle (MIP) in the thin substrate (as compared to a standard 300 $\mu$m thick silicon). Keeping the gain value at a moderate level, i.e. considerably below the avalanche regime, is instrumental  in keeping the noise low~\cite{DALLABETTA2015154}.

 In general, the time resolution of a detector system includes a few terms, which account for the noise of the detector ("jitter"), the fluctuations in the ionization density and the resolution of the time-to-digital converter.
To minimize the fluctuations in the ionization density, which ultimately set the minimum achievable time resolution, thin substrates are preferred. The time resolution  $\sigma_t$ due to the jitter is given by (see Ref.~\cite{spieler}, page 35):
\[\sigma_t=\frac{\sigma_n}{\frac{dS}{dt}}\]
where $\sigma_n$ is the  r.m.s. voltage noise of the system and $S$ the voltage signal  amplitude, thus $\frac{dS}{dt}$ is the slope of the signal ($t$ being the time). To minimize  $\sigma_t$, low-noise and  fast signals are needed. In an LGAD, the initial part of the signal produced by a MIP is due to the drift of the electrons and the signal amplitude is independent of  the gain and substrate thickness. At later times, however, the signal is dominated by the drift of the holes produced in the gain layer and it is faster in thin substrates, while its amplitude is proportional to the gain. Therefore,  thin substrates  in the order of few tens of micrometers (30-50 ${\rm \mu m}$) are typically chosen for LGAD fabrications. 

For a stable and reproducible amplification, the electric field lines inside the device, when depleted, must be straight and parallel to each other as much as possible. To achieve this goal, edge effects are minimized by designing pads  with sizes  far larger than the silicon active thickness.

At the edge of the $n^+$ implant, an additional deep $n$ implant is inserted to protect  from an early breakdown~\cite{FERNANDEZMARTINEZ201693}. 
This is usually called {\it Junction Termination Edge (JTE)}. To keep the dead area small at the edge of the devices, the JTE must be as narrow as possible. 

The LGAD, especially after irradiation,  operates at relatively high bias voltage. Consequently a Guard Ring (GR) termination  is added at the periphery of the active  area of the device. This will gradually bring the high voltage at the cutline to the ground applied to the innermost GR surrounding the pad, therefore preventing breakdowns.

 In particular, high electric fields may develop at the edges of the junctions: for a given geometry, deep  diffused implants  are expected to develop lower electric fields than shallow implants~(see Ref.~\cite{ghandhi}, page 53).  Therefore, benefiting from the JTE implantation in the process flow, the GRs are designed with this deep implant. The design of the GRs can be very similar to one of standard silicon sensor design and they can sustain similar voltages.

Several groups around the world are working on the development of the LGAD (see Ref.~\cite{1748-0221-13-03-C03014,Pellegrini:2014lki,DALLABETTA2015154}). 
This paper details the specific fabrication technology developed at Brookhaven National Laboratory (BNL) that targets the detection of MIPs. This paper is structured as follows. In Section 2 a detailed description of the fabrication process is given; in Section 3 the electrical characterization of small pads is presented, while the results of the measurements to characterize the gain performance are the subject of Section 4.

\section{Fabrication Process}
In the following we describe the process developed for the fabrication of the first LGAD prototypes at BNL. The process consists of 7 lithographies  and 4 ion-implantations (two of them at high energy). It was  carried out with wet etches in the class-100 clean room of the Instrumentation Division at BNL. 

\begin{figure}
\centering
\includegraphics[width=0.45\textwidth]{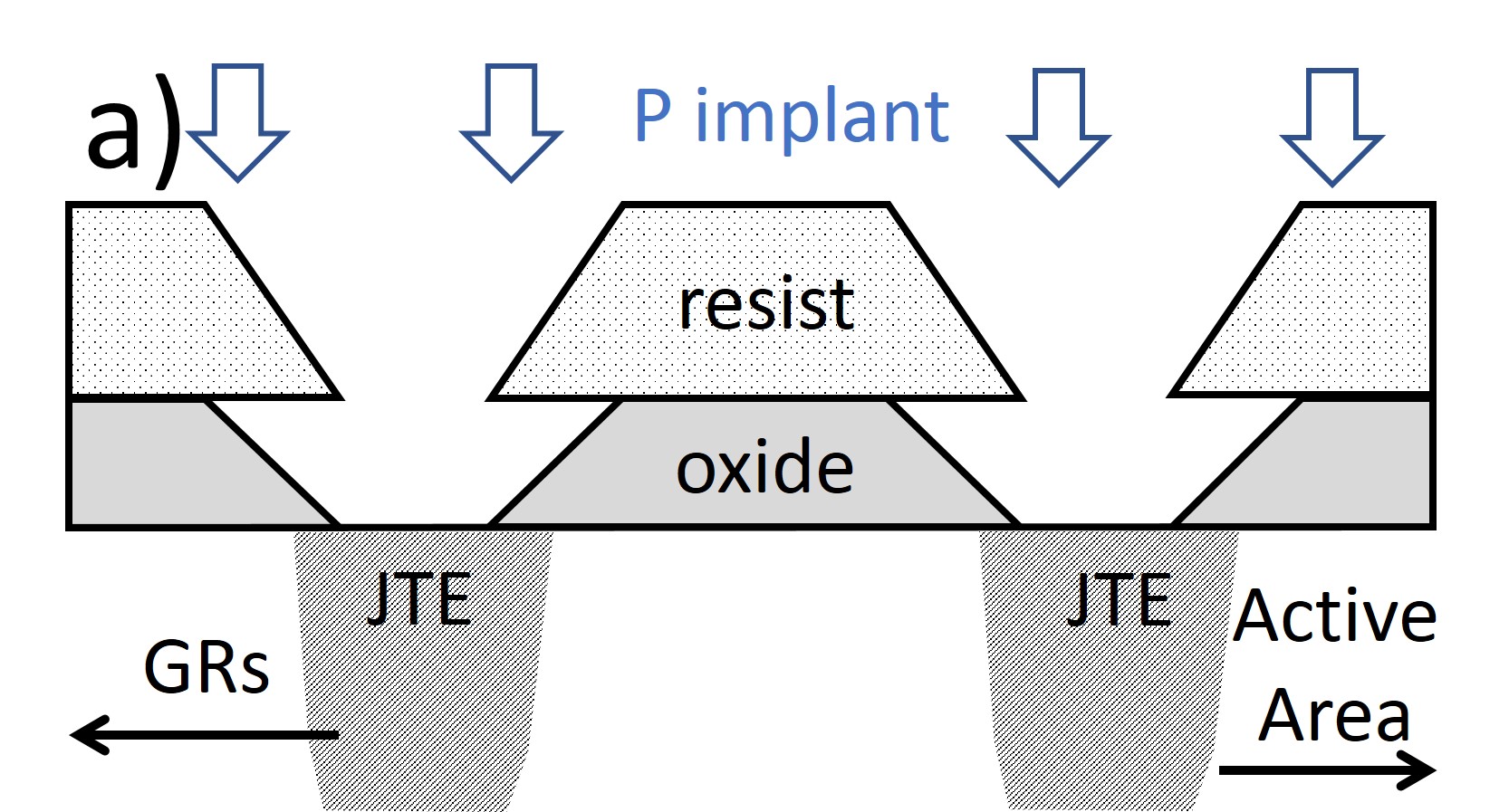}
\includegraphics[width=0.45\textwidth]{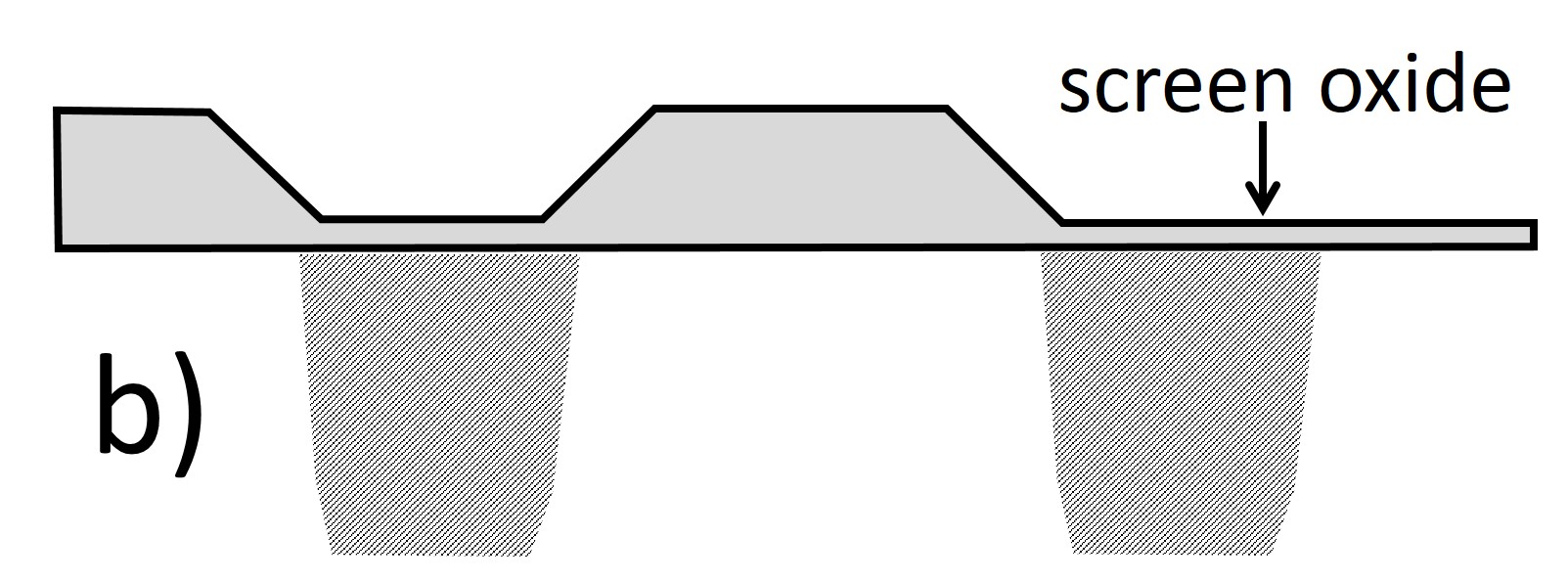}
\includegraphics[width=0.45\textwidth]{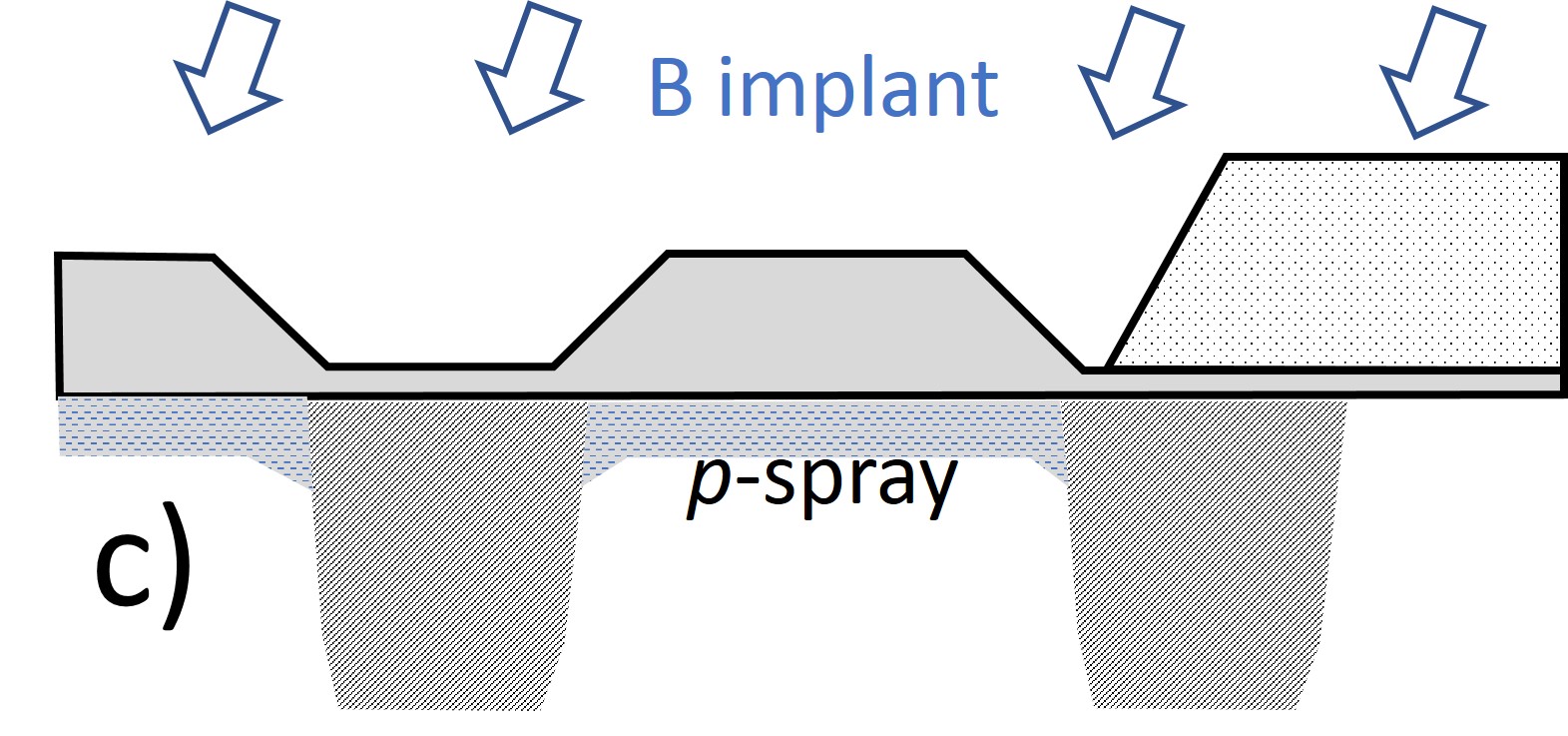}
\includegraphics[width=0.45\textwidth]{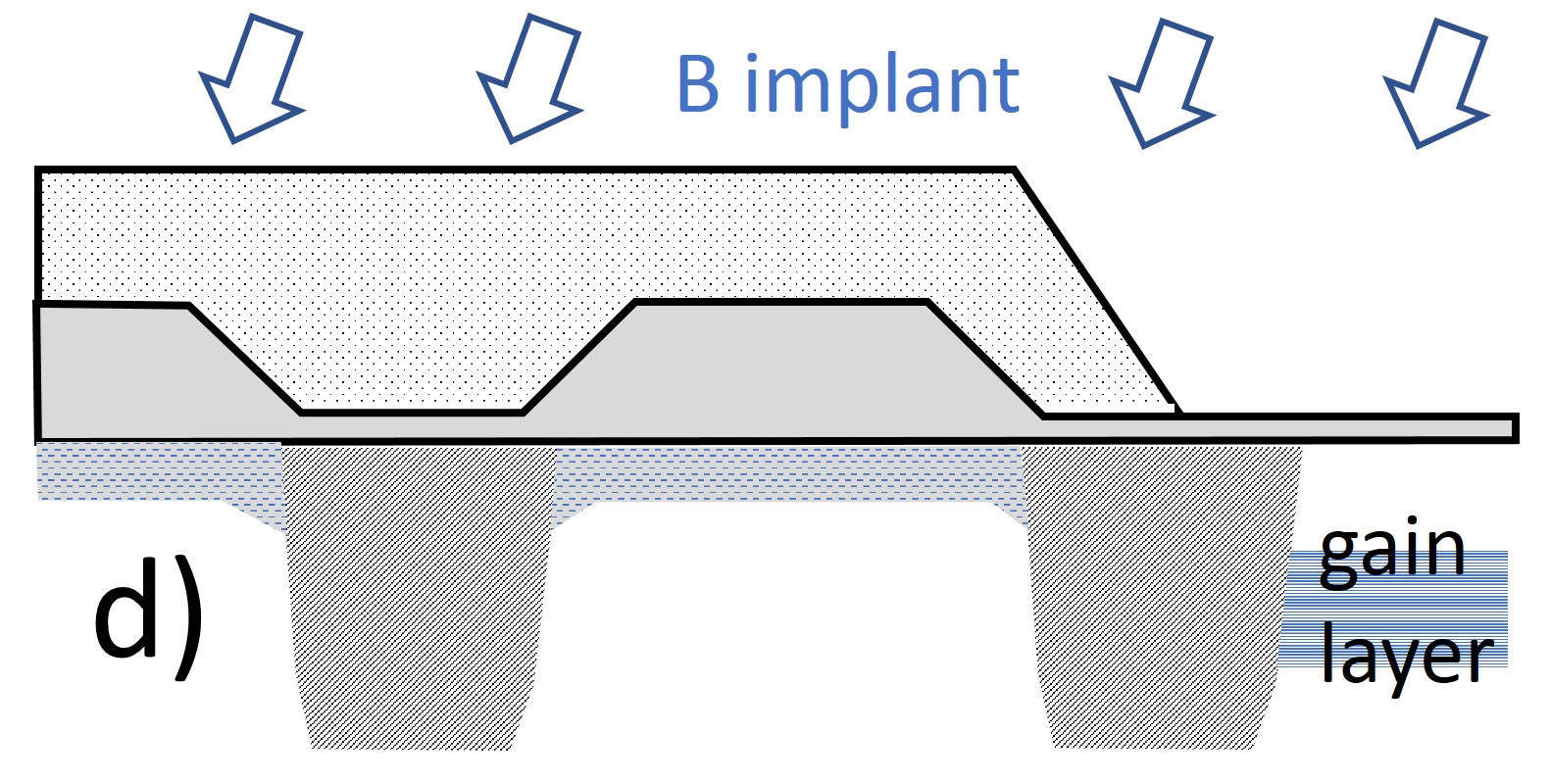}
\includegraphics[width=0.45\textwidth]{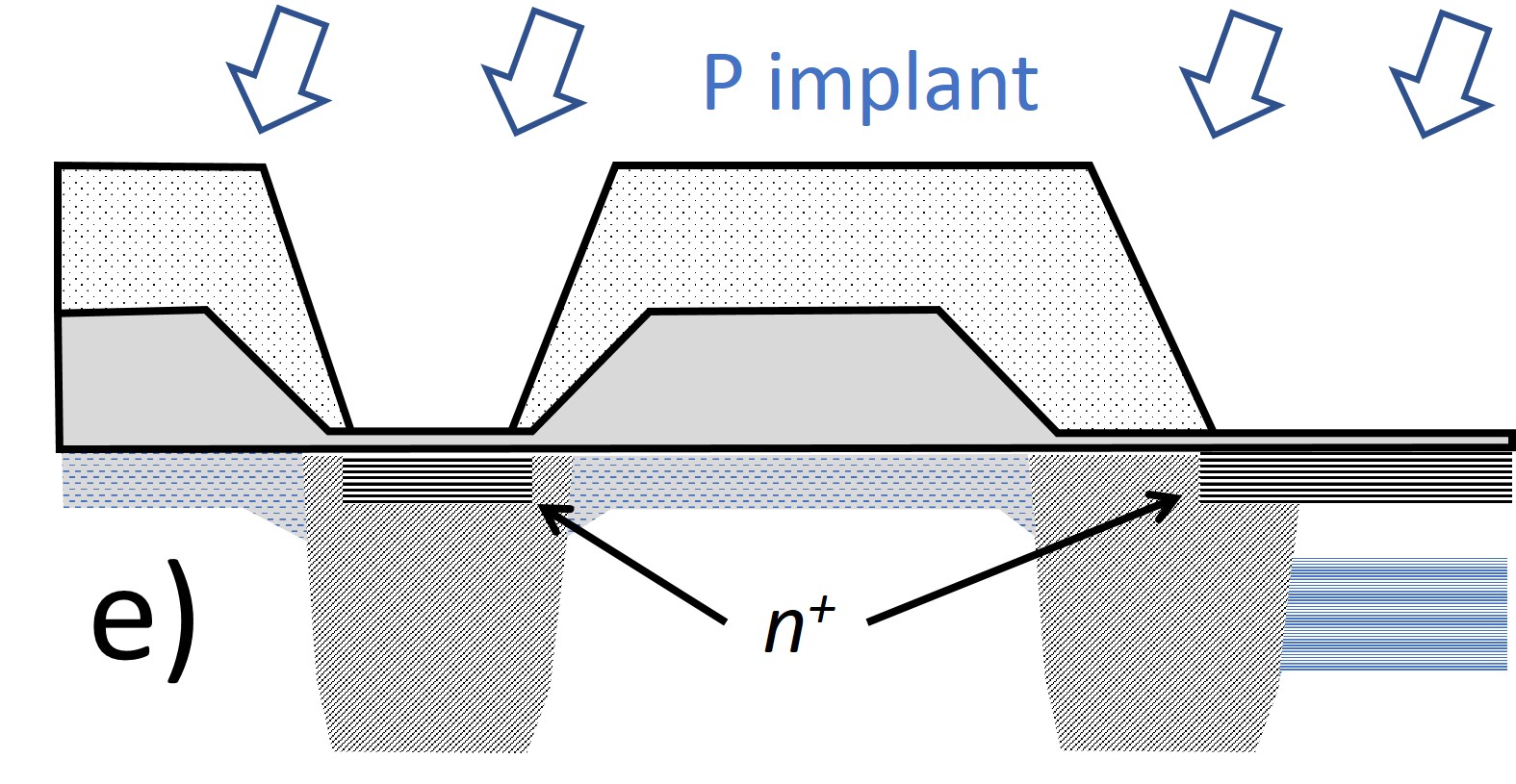}
\includegraphics[width=0.45\textwidth]{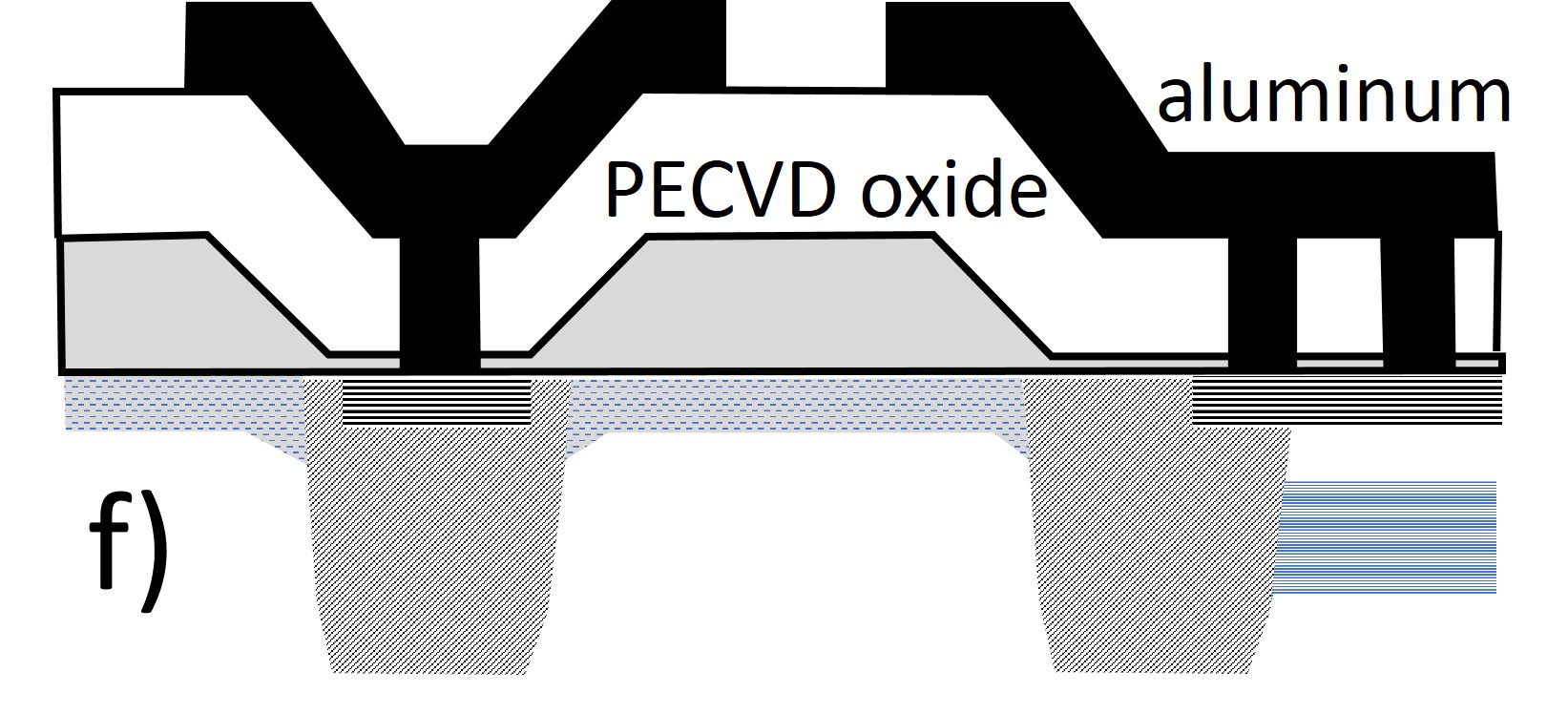}
\caption{\label{fig:process} Main steps of the LGAD fabrication process used at BNL (not to scale).  a)  JTE  definition and phosphorus (P) implantation, b) thin-layer re-oxidation, c) $p$-spray boron (B) implantation, d) gain layer boron implantation, e)  $n^+$ phosphorus implantation and  f) final device, after PECVD oxide growth, contact opening and aluminum sputtering and definition. The arrows indicate the implantation beam and its general direction. }
\end{figure}

\begin{figure}
\centering
\includegraphics[width=0.8\textwidth]{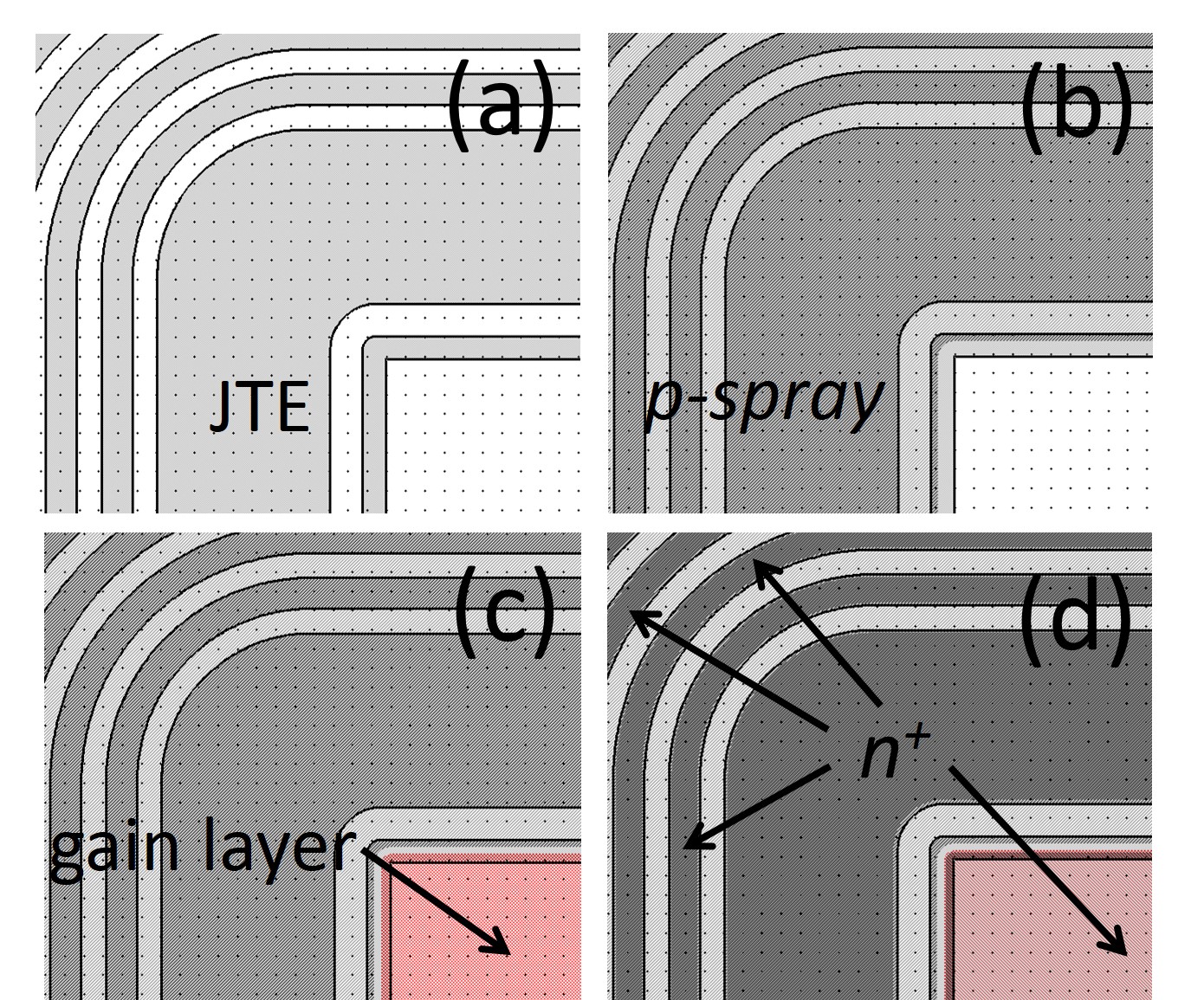}
\caption{\label{fig:layout} Layouts of corners of LGAD pads, showing various implantation regions: a) shows the JTE structure, b) the $p$-spray, c) the gain layer and finally d) the $n^+$ implantation. Grid points are $10 ~{\rm  \mu m}$ apart.}
\end{figure}

\begin{figure}
\centering
\includegraphics[width=0.8\textwidth]{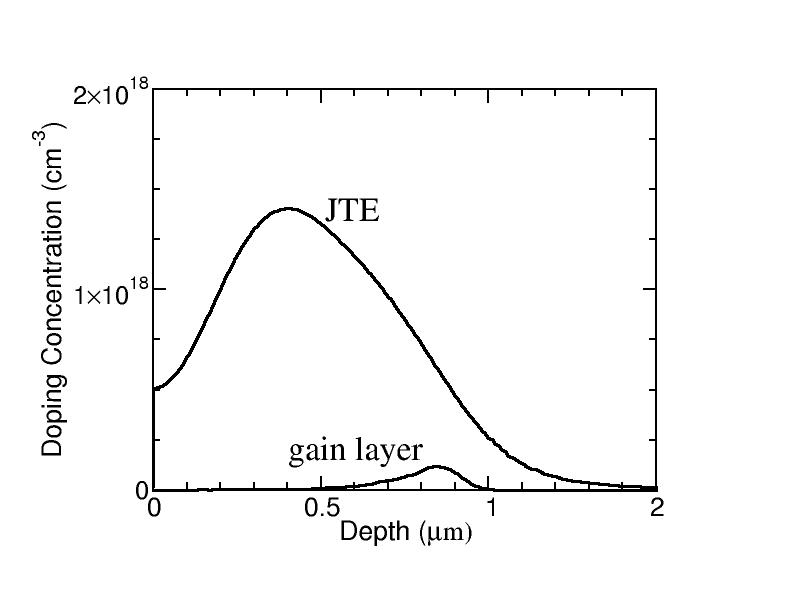}
\caption{\label{fig:simulatedprofiles} Simulated doping concentration as a function of depth for the JTE structure and the gain layer for an implantation dose of $3\cdot10^{12}~  {\rm cm^{-3}}$.}
\end{figure}

The starting material is a (100)-oriented, 4" epitaxial silicon substrate. The epitaxial layer is $p-$type, 50 $\mu$m thick with a doping concentration of about $5\cdot10^{13} ~{\rm cm^{-3}}$. The substrate on which the epitaxial layer is grown is  a heavily doped, 500 $\mu$m thick, Czochralski substrate that acts as an ohmic contact and mechanical support. 
The wafers, after the initial oxidation, undergo a series of steps, as outlined below: 
\begin{itemize}

\item The first lithographic step is the definition of the JTE (Figure~\ref{fig:process}a and Figure~\ref{fig:layout}a). The silicon oxide is removed in the areas freed from  photoresist with a high-energy phosphorus implantation following. The ion implantation is performed at zero degrees of tilt and 45 degrees of twist, to exploit the channelling  of ions in the (100) substrate.  This allows the implant to penetrate deeper  into the silicon crystal without requiring the use of higher implantation energies. On the contrary, all other implantations are performed at 7 degrees of tilt and 23 of twist to minimize channeling. A SILVACO TCAD ~\cite{silvaco} simulation is performed to predict the doping profile of the JTE and it is reported in Figure~\ref{fig:simulatedprofiles}.

\item After an additional lithography, the silicon oxide in the active area of the LGAD is removed by  wet etching. A re-oxidation follows: a thin screen oxide of about 35 nm is grown, followed by a drive-in done in nitrogen (Figure~\ref{fig:process}b).

\item 
The electron accumulation layer at the silicon to silicon-oxide interface induced by the positive charge into the oxide would short the electrodes together, therefore  a $p-$type (boron) implant (a patterned "$p-$stop" or an unpatterned "$p-$spray")  is placed in the gaps between them. In this type of fabrication of LGADs, to insulate the GRs from each other and from the LGAD pad, a  $p-$spray implant is chosen  because of its superior performances after irradiation~\cite{piemontepspray}. Differing from a standard $p-$spray, which is  uniformly implanted  through the oxide at the beginning of the process, in this production it is patterned so that this boron implant does not interfere with the gain layer in the active area which is also made of boron (Figure~\ref{fig:process}c and Figure~\ref{fig:layout}b). These   ion implantation parameters are standard for all $n$-on-$p$ processes at BNL: a dose of $2\cdot10^{12}~ {\rm cm^{-2}}$ at 100 keV,  through a 300 nm thick oxide.  The $p-$spray is implanted  in the GR regions too, where only the thin screen oxide covers the $n-$implanted silicon. However, since the dose  introduced by the $p-$spray is much lower than the GRs and its energy is low enough to keep it within the GR, the $p-$spray is fully compensated by the GR implant.

\item 

In the active LGAD area, the gain layer is implanted. For these prototypes, the energy is chosen so that the implant is not deeper than the JTE (Figure~\ref{fig:process}d and Figure~\ref{fig:layout}c). This is visible in Figure \ref{fig:simulatedprofiles}, where the JTE and the doping concentration of the gain layer are compared. Doses from $2\cdot10^{12}~ {\rm cm^{-2}}$ to $3.5\cdot10^{12}~ {\rm cm^{-2}}$ have been used to study different gain values. An annealing in nitrogen  follows.

\item 

After $n^+$ lithography and photoresist exposure, before the $n^+$ implant, a short wet etching thins down the screen oxide. Subsequently, a high-dose phosphorus implant is performed to make a good ohmic contact (Figure~\ref{fig:process}e and Figure~\ref{fig:layout}d).
A final annealing in nitrogen for the activation of the phosphorus implant  is the last thermal cycle of the fabrication.

\item 

A uniform  PECVD oxide is grown, then contacts are opened, an aluminum (1\%~silicon) layer is sputtered on both sides of the wafers and patterned on the front side (Figure~\ref{fig:process}f). Sintering in forming gas ends the process and no passivation is used for this production.

\end{itemize}

\section{Electrical Characterization}

\begin{figure}
\centering
\includegraphics[width=0.7\textwidth]{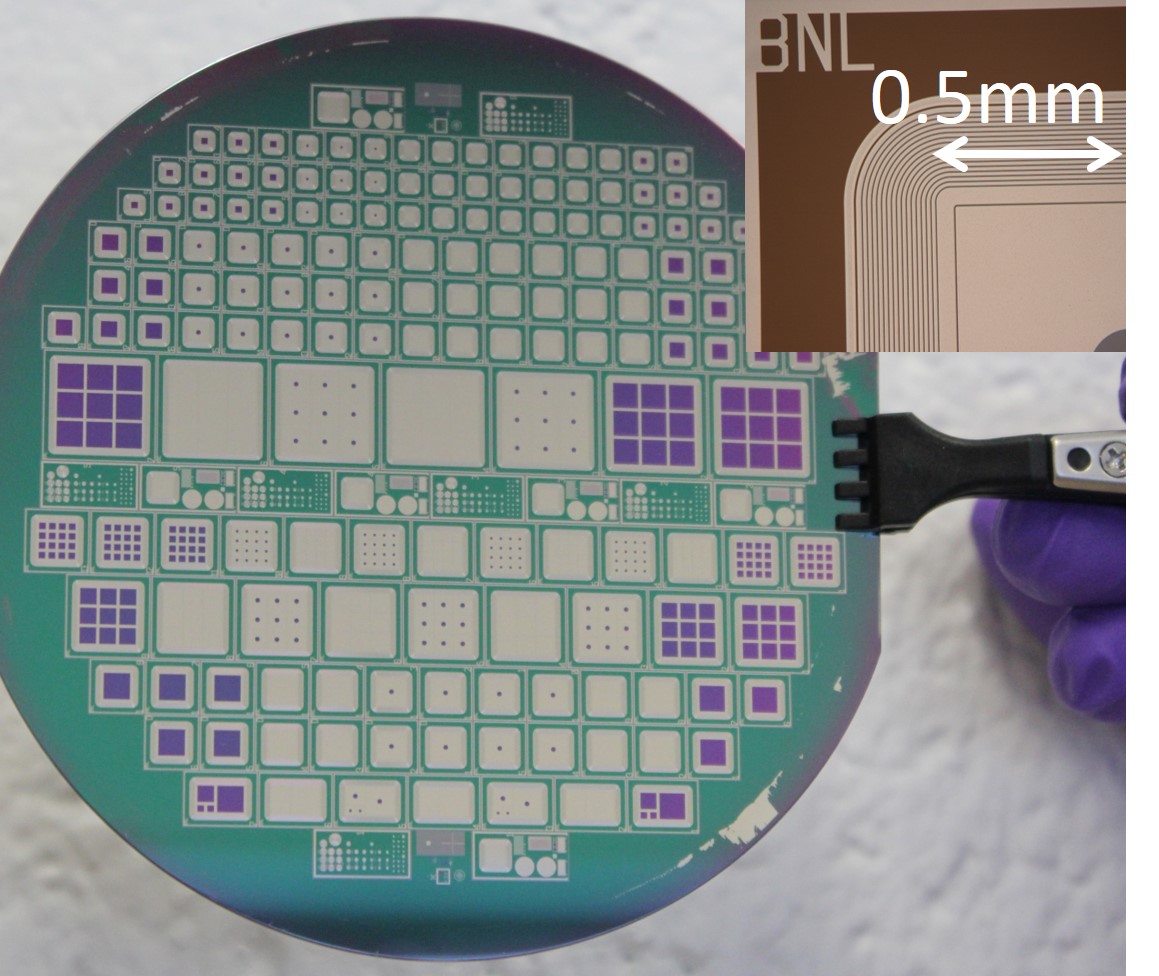}
\caption{\label{fig:wafer} Photograph of a 4" wafer populated with single LGAD pads and arrays of LGAD pads. The inset on the top-right corner shows a close-up of an LGAD structure.}
\end{figure}

A picture of a fabricated wafer  is shown in Figure~\ref{fig:wafer}. It is populated with  single-pad LGADs of 1x1 ${\rm mm^2}$, 2x2 ${\rm mm^2}$, and 3x3 ${\rm mm^2}$ with the same termination structures. 
Arrays of   pads with these dimensions are also present. Diodes which share the same dimensions and differ from the LGADs only by the absence of the gain layer are also included for reference. Standard test structures for the assessment of the quality of the fabrication batch (and the reproducibility from batch to batch) are inserted. From these, a bulk leakage current of about 1 ${\rm nA/cm^2}$ is measured as well as good electrical properties such as contact resistances, sheet resistances etc.

\begin{figure}
\centering
\includegraphics[width=0.80\textwidth]{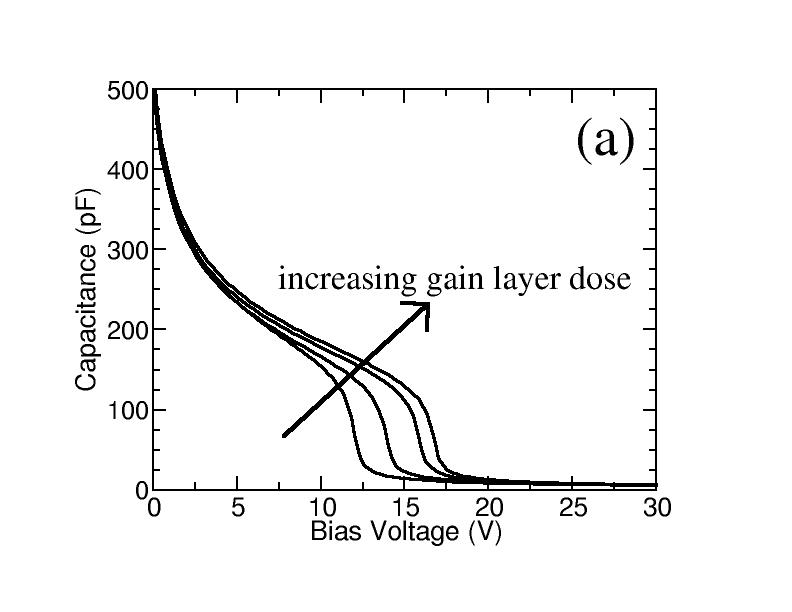}
\includegraphics[width=0.80\textwidth]{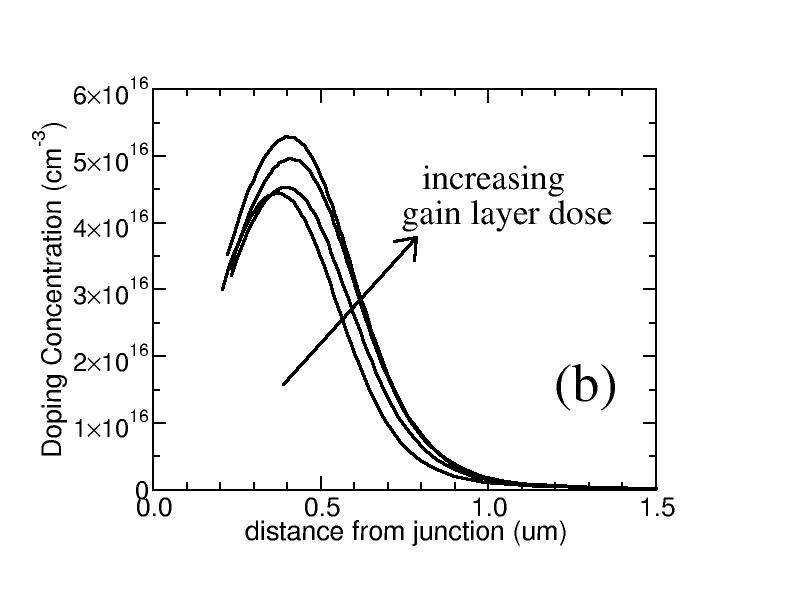}
\caption{\label{fig:CV} Top: Capacitance as a function of bias voltage for 1x1 ${\rm mm^2}$ LGAD pads. The various curves are for LGAD pads fabricated with different doping doses of the gain layer, spanning from $2.5\cdot10^{12}~ {\rm cm^{-2}}$ up to $3.25\cdot10^{12}~ {\rm cm^{-2}}$, in steps of $0.25\cdot10^{12}~ {\rm cm^{-2}}$ ($f={\rm 10~ kHz}$). Bottom: Doping concentration of the gain layer as a function of the distance from the junction, as extracted from the C-V scans on the top. The diagonal arrows point towards the direction of increasing gain layer doses.
}
\end{figure}

A critical parameter is the gain, which is controlled in this fabrication by the boron dose of the gain layer, since the implantation energy is kept fixed. For optimizing the gain, different wafers were processed at different boron dose values, starting from $2.5\cdot10^{12}~ {\rm cm^{-2}}$ up to $3.5\cdot10^{12}~ {\rm cm^{-2}}$, in steps of $0.25\cdot10^{12}~ {\rm cm^{-2}}$. The doping profile of the activated gain layer (as well as the doping of the epitaxial  layer)  can be extracted by a capacitance-voltage (C-V) measurement performed on a single pad. Throughout the measurements the GR is kept at ground voltage. Since the pad dimensions are far larger than the substrate thickness, we can obtain an accurate 1-D profile, i.e. not distorted by edge effects. The C-V scans acquired for 1x1 ${\rm mm^2}$ pads belonging to different wafers are reported in Figure~\ref{fig:CV}. The high capacitance at the lowest bias voltages is due to the incomplete depletion of the gain layer. 
As soon as the gain layer is depleted and the depletion region starts extending into the much less doped epitaxial layer, the capacitance abruptly drops, creating a {\it foot} in the C-V curve. For increasing values of the gain layer dose, the foot shifts to higher voltage values.

By applying the following formula (see Ref.~\cite{grove}, page 171)
\[N_{c} (x)= -\frac{C(V)^3}{\epsilon_{si}\cdot q\cdot A^2 \cdot\frac{dC(V)}{dV}}~~,\]
with \(x=\frac{A\cdot \epsilon_{si}}{C(V)}\) being the depleted thickness at the voltage $V$, $A$ the area of the junction, $\epsilon_{si}$ the dielectric constant of silicon, and $q$ the elementary electrical charge, one can extract the doping concentration $N_c$ of the gain layer as a function of the depth $x$, starting from the metallurgical junction between the $n^+$ and the gain layer itself. If we assume that the distribution of doping concentration is symmetric with respect to its peak in $x$, it is apparent from Fig.~\ref{fig:CV} that part of the gain layer is compensated by the $n^+$ layer, and thus a fraction of the boron dose is not effective in creating the electric field that produces multiplication. 
Furthermore, because the gain layer is relatively shallow, only the deepest part of the doping profile can be extracted from the results of C-V scan. Numerically integrating in $x$ the distribution of doping concentration from the position of the peak to the  substrate, and doubling the result, we can infer the  fraction of dopant that is electrically active,  in the hypothesis that the doping profile is symmetric with respect to the peak. Table~\ref{table:table} reports the values of the dose extracted from the C-V scan, the inferred fractions of active dose and the measured breakdown voltage for each value of the nominal implantation dose. The table shows that  activation fractions in excess of 0.9 are obtained.  However, these values must be considered as a lower limit, because implantations are retrograde  and thus not symmetric: the dose in the silicon on the left of the peak is (slightly) larger than the dose on the right of the peak along the depth $x$. This can be appreciated in the example of the simulated profiles reported in Figure \ref{fig:simulatedprofiles}.

\begin{figure}
\centering
\includegraphics[width=0.9\textwidth]{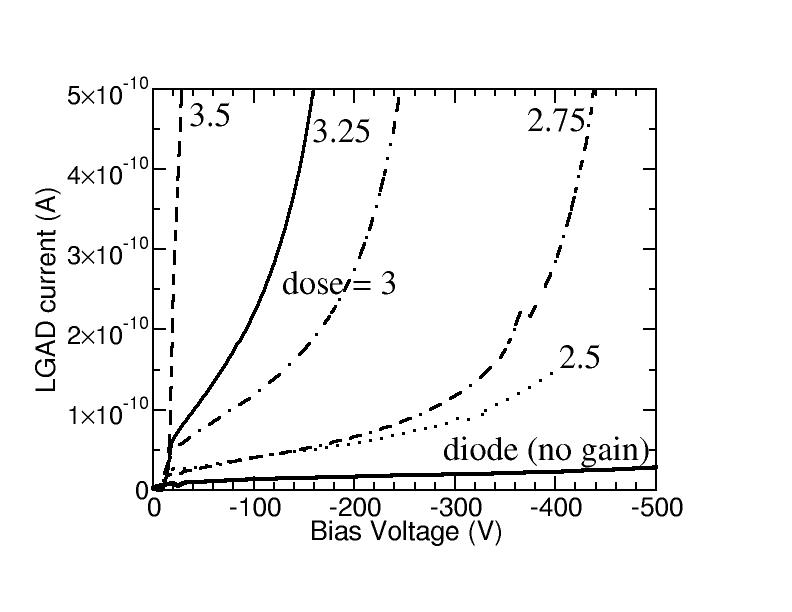}
\caption{\label{fig:IV} Examples of I-V characteristics for 1x1 ${\rm mm^2}$ LGAD pads fabricated in wafers implanted with different gain layer doses (in units of $10^{12}~ {\rm cm^{2}}$). For reference the current of a 1x1 ${\rm mm^2}$ diode, i.e. with no gain, is shown. In these measurements the GR is grounded.}
\end{figure}

Samples of the current-voltage characteristics (I-V), as measured on 1x1 ${\rm mm^2}$ LGADs belonging to wafers implanted with different gain layer doses, are shown in  Figure~\ref{fig:IV}. The gain layer implantation dose strongly affects the breakdown voltage (reported in Table \ref{table:table}), and the amount of the pad current, which is merely the leakage current augmented by the gain.
However, other methods are employed to measure the gain: a Charge Sensitive Pre-Amplifier (CSA) was used, as detailed in the following section.
 
 \begin{table}
\begin{tabular}{|c|c|c|c|}
\hline
 Nom. Dose ($10^{12}~ {\rm cm^{-2}}$)  & Dose from C-V ($10^{12} ~{\rm cm^{-2}}$) & Activation Frac. & ${\rm V_{BD} ~(V)}$    \\
 \hline
 2.50  & 2.3 & 0.92 & 450   \\
 2.75 & 2.5 & 0.91 & 380   \\
 3.00    & 2.7 &  0.90 & 300 \\ 
 3.25 & 2.9 &  0.89 & 200   \\
 3.50  & 3.2 &  0.91 & 50  \\
 \hline
\end{tabular}
 \caption{For each value of the nominal implantation dose, the values of the dose extracted from C-V scan, the activation fraction and the breakdown voltage (${\rm V_{BD}}$) are reported.}
  \label{table:table}
\end{table}
 
\section{Gain Measurements}
A few 3x3 ${\rm mm^2}$ LGAD pads were cut out from various wafers and assembled on  test-boards by gluing and wire-bonding in order to make an electrical connection to the input of the CSA. For calibration purposes, diodes fabricated on the same wafers were mounted on the same test-boards. Diodes differ from the LGADs  only in the absence of the gain layer, while they share all other properties such as dimensions and, thus, capacitance (18~pF at full depletion regime). The CSA is a single-channel model, developed at BNL, whose input capacitance is in the order of 20~pF. The output signal is fed to an AMPTEK PX5 ~\cite{px5}, for spectra acquisition. Finally, either a gamma ray ${}^{241}$Am or an X-ray  ${}^{55}$Fe source shines over the devices.

The former has the most energetic line at 60 keV  generating about $16.6\cdot 10^3$ electron/hole pairs ($e^-/h^+$) in silicon, while the latter has a line at 6 keV which thus creates one tenth of the charge and about half of the average number of $e^-/h^+$  generated by a MIP in 50  $\mu$m of silicon ( $4\cdot 10^3$ $e^-/h^+$). 

\begin{figure}
\centering
\includegraphics[width=0.9\textwidth]{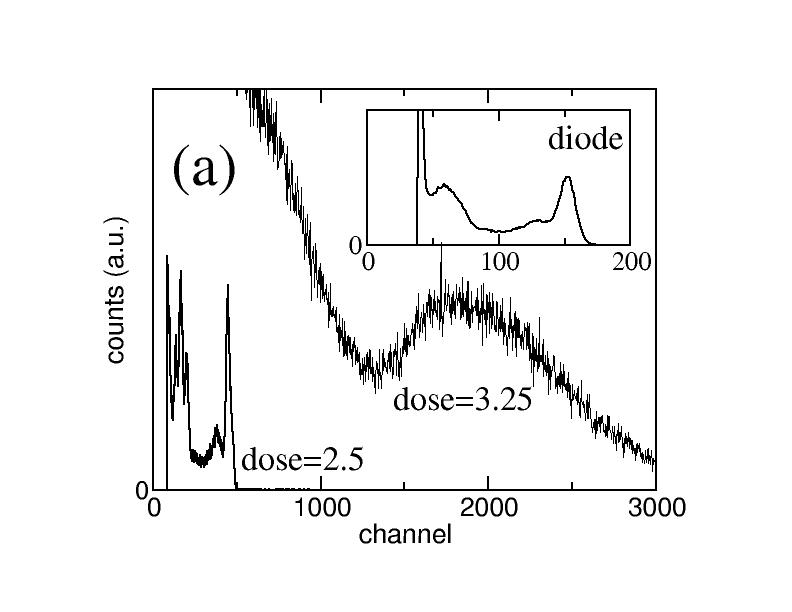}
\includegraphics[width=0.9\textwidth]{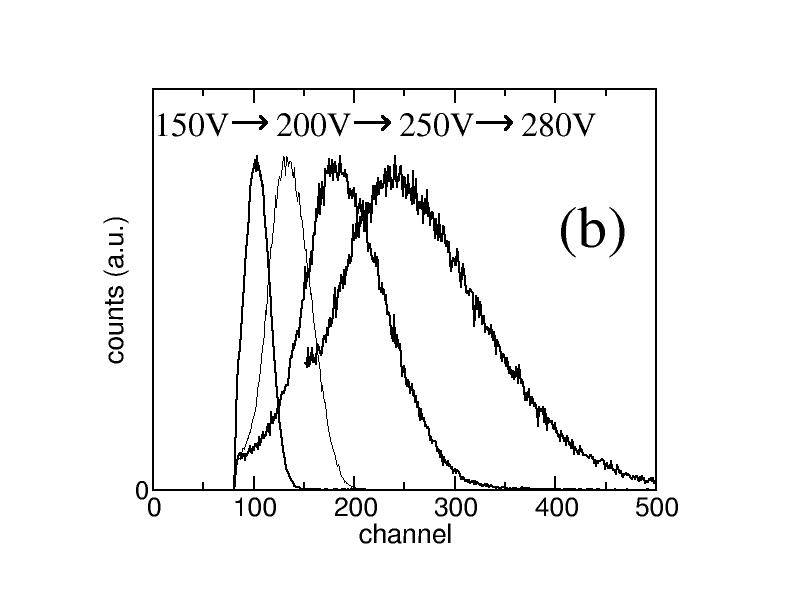}
\caption{\label{fig:spectra} (a) The ${}^{241}$Am spectra acquired with LGADs featuring a gain layer dose of either $2.5\cdot10^{12} ~{\rm cm^{2}}$ or $3.25\cdot10^{12} ~{\rm cm^{2}}$. In the inset, the same spectrum acquired by a diode (gain=1). (b) The ${}^{55}$Fe spectra aquired with LGADs featuring a gain layer dose of $3\cdot10^{12} ~{\rm cm^{2}}$, for different bias voltages. The $x$-axis shows the PX5 ADC channel number, the $y$-axis shows the counts per channel. The counts for each spectrum are normalized for an easier comparison of the spectra.}
\end{figure}

Examples of spectra acquired with this set-up are shown in Figure~\ref{fig:spectra}. As compared to the spectrum acquired with the diode, for low implantation doses in the gain layer (thus low gain), the signal-to-noise ratio increases and the low-energy peaks from the ${}^{241}$Am gamma ray spectrum are visible. On the contrary, by increasing the implantation dose (thus high gain), the 60~keV peak of the spectrum clearly moves to higher channel numbers and it becomes far broader, while the low-energy peaks become indistinguishable.

In this set-up, the ${}^{55}$Fe peak is invisible if acquired either with the diode or at low gains and can only be seen at the high gains. Figure~\ref{fig:spectra} shows the ${}^{55}$Fe spectrum for a dose of $3\cdot10^{12} ~{\rm cm^{2}}$ (i.e. high gain), for different bias voltages. 

\begin{figure}
\centering
\includegraphics[width=0.85\textwidth]{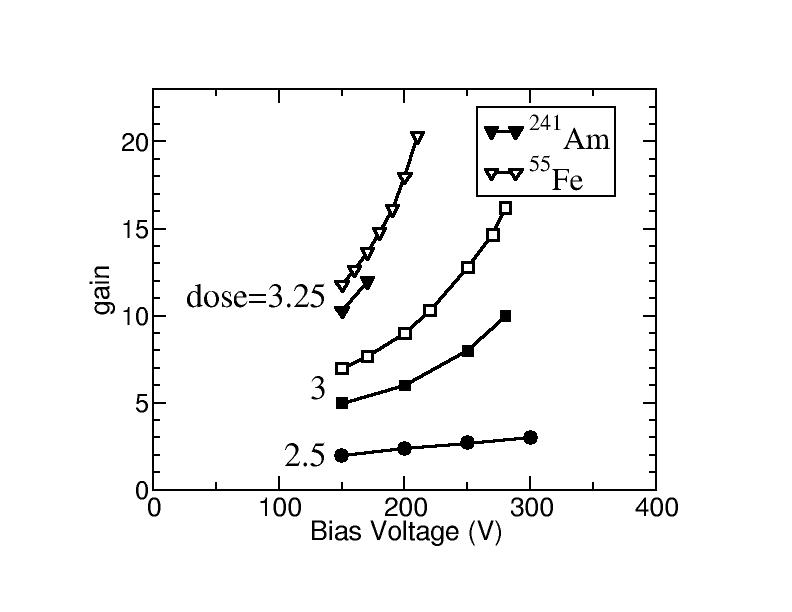}
\caption{\label{fig:gain} Gain values measured from  either the ${}^{241}$Am 60~keV (solid markers)  or the ${}^{55}$Fe 6~keV (empty markers) peak positions. Measurements are shown for three LGADs differing for the values of the implant doses of their gain layer (in units of $10^{12} ~{\rm cm^{2}}$).}
\end{figure}

The 60~keV and the 6~keV peak positions are acquired for different values of the bias voltage and are compared to the peak position using the diode as a detector. The ratio between  the peak positions for the cases of the LGAD and the diode is a measure of the LGAD gain. A summary of the LGAD gain values acquired for different bias voltages is shown in Figure~\ref{fig:gain}. Figure~\ref{fig:gain} shows that the gain is higher when measured by using the ${}^{55}$Fe source,   likely due to a charge self-shielding effect that comes into play for high values of generated charge.
While this production spans a wide range of gains, a maximum gain of 20 is observed. 


\begin{figure}
\centering
\includegraphics[width=0.85\textwidth]{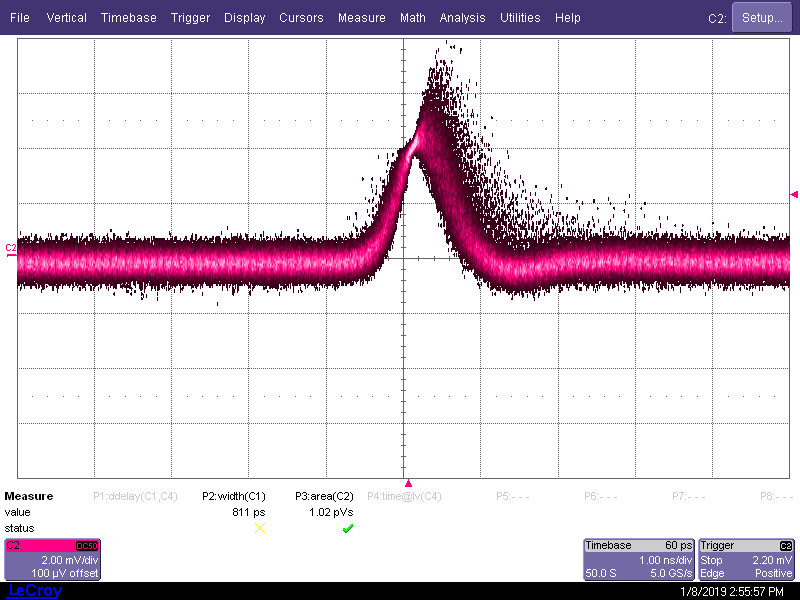}
\caption{\label{fig:waveform} Oscilloscope screenshot of signals generated by the 6~keV X-rays from an ${}^{55}$Fe radioactive source, shining over  a 1x1 ${\rm mm^2}$ LGAD and read out by a fast TA. The gain layer dose of this LGAD is $3\cdot10^{12} ~{\rm cm^{2}}$; $V_{bias}=-260$ V. Horizontal scale is 1 ns/division, vertical scale is 2 mV/division.}
\end{figure}

 A few LGADs were mounted on different test-boards and connected to a single-channel fast transimpedance amplifier (TA) developed by the Santa Cruz Institute for Particle Physics (SCIPP) at the University of California at Santa Cruz ~\cite{cartigliabeamtest}. Signals generated by X-rays from an ${}^{55}$Fe radioactive source were acquired by a 1 GHz oscilloscope. The results are shown in Figure~\ref{fig:waveform}: signals are short and develop in less than 2 ns.

\section{Conclusions}
Fast-timing detectors have attracted world-wide interest for applications in high-energy collider physics as well as in photon science and imaging. Specifically, silicon-based detectors with internal gain have made significant progress in recent years. The technology for the fabrication of Low Gain Avalanche Detectors has been developed at BNL for the detection of minimum ionizing particles. This paper details the steps of the specific fabrication process developed at BNL. The prototypes show low leakage currents, good electrical characteristics and, most importantly, gain values in a range of 2 to 20. Further optimization of the fabrication process could increase the gain to even higher values.

\section{Acknowledgements}
The authors are indebted to Dr. Marco Bomben of LPNHE,  Paris, France, for the help in the TCAD numerical simulations. The authors also want to thank Enrico Rossi (student in Master's in Instrumentation at Stony Brook University, New York, U.S.A.) as well as Alexander Langedijk and Maris Serzans (undergraduate students at University of Oxford, U.K.) for their help with the development and testing of programs for silicon sensor characterization. 
This material is based upon work supported by the U.S. Department of Energy under grant DE-SC0012704.

\bibliographystyle{elsarticle-num}
\bibliography{biblio}{}

\end{document}